\documentstyle[preprint,aps,amstex,floats,tighten]{revtex}
\input epsf
\begin{document}
\pagestyle{myheadings}

\title{Discrete energy spectrum
of Hawking radiation from Schwarzschild surfaces}

\author {Ashutosh V. Kotwal \\ {\em Department of Physics, Duke University, Durham, NC 27708-0305, USA} \\ 
\vspace*{0.2in}
Stefan Hofmann \\ {\em Institut f\"{u}r Theoretische Physik, J. W. Goethe Universit\"{a}t, 60054 Frankfurt
 am Main, Germany} }
\maketitle
\vspace*{-0.1in}
\begin{abstract}
We analyze the allowed energy levels 
of Hawking radiation from Schwarzschild surfaces
in space-times with large extra dimensions.
From the requirement that the wave functions
associated with these particles
be single-valued and obey radial
 boundary conditions, we derive 
an upper bound of their discrete energy spectrum
as a function of the number of large extra
dimensions and the mass of the black hole.
Furthermore, we investigate the spacing
of the energy levels.
\end{abstract}


Space-times with large extra dimensions (LXDs)
have been proposed \cite{add,rs} to explain or at
least reformulate the hierarchy between the
electroweak scale $M_{\rm w}\approx 10^3$ GeV 
and the Planck scale $M_{\rm Pl}\approx 10^{19}$ GeV 
in four dimensions. In these brane world scenarios the hierarchy
is either generated through the volume of the LXDs \cite{add}
or through a exponential warp factor \cite{rs} taking into account
the energy density on the brane. A common feature of these
scenarios is that all standard model particles 
including the gauge degrees of freedom 
are localized on a $3-$brane whereas gravity
propagates in both the compact LXDs and 
the non-compact dimensions.
As a consequence the fundamental scale of gravity $M_{\rm f}$
could be as low as $M_{\rm w}$ in these space-times.

The experimental bound on $M_{\rm f}$ from the absence of
missing energy signatures is $M_{\rm f}\ge 800$ GeV
\cite{wells,peskin,peskin2}.
Therefore, LXDs may be discovered at Tevatron \cite{bleicher}
and further explored with future colliders like 
LHC, TESLA or CLIC.
One possible phenomenon could be the formation
of micro black holes in high energy collider
experiments \cite{land} or atmospheric ultra high energy
cosmic rays \cite{ring}. 
The production of micro black holes in the
final state of a high energy collision
could be observed due to a sharp cut-off \cite{sqm}
in $\sigma (p\bar{p}\rightarrow {\rm Jet+X})(P_\bot)$
at Tevatron or $\sigma (p p\rightarrow {\rm Jet+X})(P_\bot)$
at LHC. 

One of the most remarkable relationships in theoretical physics
is that between classical black hole physics and the ordinary laws
of thermodynamics \cite{bardeen}. 
This relation is a striking
mathematical analogy and suggests a physical connection.
The quantity in classical black hole physics which plays
the role mathematically analogous to the total Energy $U$
is the mass $M_{\rm bh}$ of the black hole, which is physically
the total energy of the black hole in general relativity.
The quantity analogous to the temperature 
is the surface gravity $\kappa$. In classical
black hole physics $\kappa$ has nothing to do with the
physical temperature of a black hole, which
is absolute zero. However, the analysis of the behavior
of a quantum field 
results
in the Hawking effect \cite{hawking}: 
A black hole will radiate exactly
like a black body with temperature $T_{\rm bh} = \kappa / 2 \pi$.
The thermal spectrum of radiated quanta can be calculated
by noting that the physical entropy $S$ of a black hole
is proportional to its Schwarzschild surface area.
For a spherically symmetric solution in $D=4+d$ dimensions
$S\sim M_{\rm f}^{\; 2+d} A_{(2+d)} R_{\rm bh}^{\;\; 2+d}$,
where $A_{(2+d)}$ denotes the $2+d$-dimensional surface
of a $3+d$-dimensional unit sphere and
the horizon radius is given by \cite{myers}
\begin{equation}
\label{hor}
R_{\rm bh}^{\; \;1+d}
=
\frac{4}{2+d}
\frac{A_{(2)}}{A_{(2+d)}}
\left(\frac{1}{M_{\rm f}}\right)^{1+d}
\frac{M_{\rm bh}}{M_{\rm f}}
\;.
\end{equation}
In the literature \cite{harms} 
it was argued that the thermal
spectrum of radiated quanta should be calculated
using the micro-canonical ensemble. 
In the following we formulate two essential conditions
which restrict this spectrum due to finite size effects.
It is important to note that no special ensemble
is considered in deriving the main result
Equation (\ref{res}).

\vspace*{0.1in}
{ \em Condition 1 }: 
The energy $\omega$ of the radiated quanta 
has an upper bound provided by the mass $M_{\rm bh}$
of the black hole
\begin{equation}
\omega \le M_{\rm bh} / 2
\; ,
\end{equation}
which is mandatory due to energy-momentum conservation.
This condition is equivalent to the statement that
the de Broglie wavelength of the quantum formed
on the Schwarzschild surface needs to be larger than 
the Compton wave length of the micro black hole.

\vspace*{0.1in}
{\em  Condition 2 }:
\begin{itemize}
\item[(a)] 
For every de Broglie wavelength $\lambda$
associated with a particle on the Schwarzschild surface
exists a positive integer $n$ with
\begin{equation}
n \lambda = R_{\rm bh}
\; ,
\end{equation}
which implies $\lambda \le R_{\rm bh}$.
\item[(b)]
The radial wave function $F$ has to vanish
on the Schwarzschild surface, i.~e.~
\begin{equation}
F(r) = 0 \hspace{0.5cm} {\mbox {\rm for}}\hspace{0.5cm}
r \sim R_{\rm bh} \; .
\end{equation} 
\end{itemize}
Condition (a) is necessary and sufficient
in order to allow for periodic boundary
conditions so that the wave-function of a particle
localized on the Schwarzschild surface is single-valued.
If the wavelength is larger than the horizon radius
destructive interference will suppress the amplitude
for such a particle state.
Condition (b) ensures that the wave function of the
particle does not penetrate beyond the horizon.
Conditions (a) and (b) are non-trivial boundary conditions
that result in a geometrical quantization of all components
of the particle momentum $\vec{k}$.

\vspace*{0.1in}
{ \em Application to black holes }:

Ignoring numerical prefactors 
we obtain from {\em Condition 1} and {\em Condition 2}
the general relation
\begin{equation}
\label{self}
\frac{1}{M_{\rm f}}
\left(\frac{M_{\rm bh}}{M_{\rm f}}\right)^{\frac{1}{1+d}}
\ge \lambda \ge \frac{1}{M_{\rm bh}}
\end{equation}
implying $M_{\rm bh}\ge M_{\rm f}$. For $d=0$
this relation is easily satisfied for astrophysical
black holes, thereby allowing a large range for $\lambda$.
However, the above stated conditions imply 
a general self-consistency relation for arbitrary $d$.

\vspace*{0.1in}
{\em Modification of black body spectrum }:

The remarkable conclusion known as Hawking effect
is that, at late times, a Schwarzschild black hole
formed by gravitational collapse radiates precisely
as a thermal black body at temperature 
$T_{\rm bh}=(1+d)/ (A_{(2)} R_{\rm bh})$.
The number of momentum modes that can
be thermally populated is roughly
$R_{\rm bh}^3 k^2 {\rm d}k$. 
If this number changes continuously in momentum space,
the amount of energy radiated away at a given $T_{\rm bh}$ 
 is an integral of the Planck spectrum 
\begin{equation}
u(\omega, T_{\rm bh})
=
\sum
\frac{g}{2\pi^2} k^2 f(\omega/T_{\rm bh}) \; \omega
\end{equation}
over the energy. Here, the sum is over all possible
particle species, $g$ counts the effective degrees of freedom
and $f$ denotes the Bose-Einstein or the Fermi-Dirac
distribution respectively for every particle species.
Planck's radiation law is reliable for
$M_{\rm bh}/T_{\rm bh} \gg 1$, where in our case
$M_{\rm bh}/T_{\rm bh} \approx$ 
$(M_{\rm bh}/M_{\rm f})^{(2+d)/(1+d)}$.

In the following we examine when the assumption 
of continuous variation of allowed momentum values
really holds.
Let us consider a Schwarzschild sphere
centered in a cube with side length $2 R_{\rm bh}$.
The energy spectrum of modes on the Schwarzschild surface 
can be approximated by the spectrum of modes
inside the cube.
Then, {\em Condition 2} is modified by
$R_{\rm bh} \rightarrow 2 R_{\rm bh}$.
As a consequence, the relation $k= \pi n / R_{\rm bh}$
between the momentum $k$ of the particle (with mass $m$), 
the positive integer $n$ and the horizon radius follows.
With {\em Condition 1} we find our main result
\begin{eqnarray}
\label{res}
n (d, M_{\rm bh})
&\le& (4\pi^3)^{-\frac{1}{2}}
\left(1-\frac{4m^2}{M^2_{\rm bh}}\right)^{\frac{1}{2}}
\nonumber \\
&&\left(\frac{8}{2+d}\;  
\Gamma\left(\frac{3+d}{2}\right)\right)^{\frac{1}{1+d}}
\left(\frac{M_{\rm bh}}{M_{\rm f}}\right)^{\frac{2+d}{1+d}}
\; .
\end{eqnarray}
Again, for $d=0$ and macroscopic black holes
we have typically $n\sim 10^{60}$
and therefore a continuous thermal spectrum
of radiated quanta.
On the other hand,
for $M_{\rm bh}\sim M_{\rm f}$, $n$ is of order one
and $k$ is quantized due to the small size of
the Schwarzschild surface. {\em Condition 2}
then restricts the phase space in the infrared
and allows no modes with $k \ll T$.
Note that in deriving (\ref{res}) 
we did not specify any ensemble.

\begin{figure}[tbhp]
\epsfxsize=3.0in
\centerline{\epsfbox{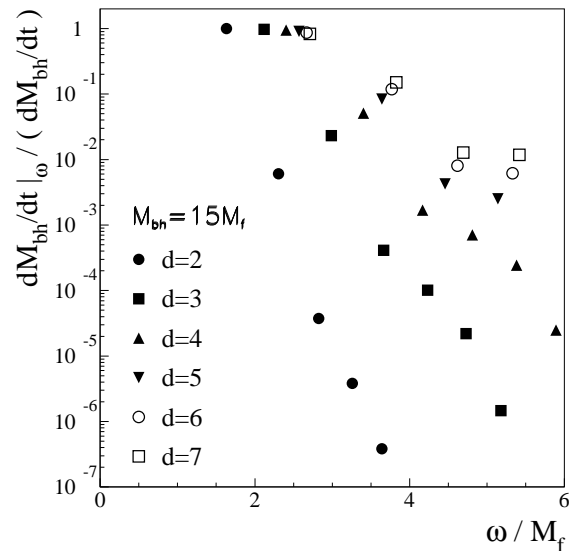}}
\caption{The power emitted via standard model particles
 in a certain frequency mode
normalized to (\protect\ref{ps})
as a function of $\omega/M_{\rm f}$
for $d\in\{2,3,4,5,6,7\}$.
}
\label{disfrac}
\end{figure}

As a consequence, for black holes with 
$M_{\rm bh}\sim M_{\rm f}$ only a discrete
energy spectrum is possible with finite spacing between the levels.
Instead of integrating the Planck spectrum
over the energy we have a sum over
the discrete levels
\begin{equation}
\label{ps}
\sum\limits_{n=1}^{[n\le n_{\rm max}]}
u(\omega_n, T_{\rm bh})
\; .
\end{equation}
Here, $n_{\rm max}$ is the right hand side of
(\ref{res}) and $[\dots]$ denote the largest
integer part of the argument in brackets. 
The energy spacing between a mode with momentum $k$ and a mode with
momentum $k+{\rm d}k$ is approximately $1/R_{\rm bh}$.
Since (\ref{res}) is general, similar conclusions
should hold for the micro-canonical ensemble, which we will discuss
 in a forthcoming publication.
Here we illustrate our results
using the grand canonical ensemble, where
\begin{equation}
\frac{{\rm d}M_{\rm bh}}{{\rm d}t} 
=
- A_{(2)} \; R_{\rm bh}^2  
\sum\limits_{n=1}^{[n\le n_{\rm max}]}
u(\omega_n, T_{\rm bh})
\end{equation}
for standard model fields \cite{emparan}, 
 taking into account finite spacing between the energy levels.
For a black hole with $M_{\rm bh} = 15 M_{\rm f}$ and $d=2$ we find
$n_{\rm max}\approx 5$. However, the spectrum is peaked
for $\omega \approx M_{\rm f}$, i.\,e.~for the lowest 
energy level accessible for a particle on the Schwarzschild
surface.
Nearly all energy goes in the lowest possible energy mode,
as shown in Fig.~\ref{disfrac}.
Neglecting modes with higher energies results in a relative error
of the order of one percent. For $d=3-7$ the situation is similar,
though $n_{\rm max}$ is decreasing and
the maximum relative error is of the order of ten percent.
With increasing $\omega$ the $d$-dependence becomes more
important. For example, for the second energy level
we find $\omega\approx 2 M_{\rm f}$ for $d=2$
and $\omega\approx 4 M_{\rm f}$ for $d=7$.

In conclusion, we have shown that the energy spectrum
of particles emitted by $M_{\rm f}$-scale black holes  
is quantized due to non-trivial boundary conditions.
For the lowest lying energy level accessible for 
particles on the Schwarzschild surface we find
$\omega\approx M_{\rm f}$. Nearly all energy 
is radiated off the Schwarzschild surface in modes 
with this lowest energy.
This may have dramatic consequences. Depending
on the detailed prescription for geometrical
quantization, the black holes on the $M_{\rm f}$-scale
may not evaporate at all via Hawking radiation. 

\acknowledgements
A.~K.~ thanks C.~Hays and
S.~H.~ thanks S.~Hossenfelder for fruitful discussions.
S.~H.~ is grateful to Steffen A.~Bass 
for his hospitality at Duke University 
and partial support through U.S. Department of Energy
  grant DE-FG02-96ER40945.
The work of A.~K.~ was supported in part by the U.S.
Department of Energy and the Alfred P. Sloan Foundation.

\end{document}